\documentclass[twocolumn,showpacs,preprintnumbers,amsmath,amssymb]{revtex4}
\usepackage{graphicx}% Include figure files
\usepackage{dcolumn}% Align table columns on decimal point
\usepackage{bm}% bold math
\def\bea {\begin{eqnarray}}
\def\eea {\end{eqnarray}}

\def\be {\begin{equation}}
\def\ee {\end{equation}}

\newcommand{\Gm}{\Gamma}

\newcommand{\del}{\partial}

\newcommand{\F}{F_\pi}
\newcommand{\D}{{\cal D}}
\newcommand{\oK}{\overline{K}}

\newcommand{\slv}{v \!\!\! /}

\begin{document}

\title{Drag and diffusion coefficients of $B$ mesons in hot hadronic matter}

\author{Santosh K Das, Sabyasachi Ghosh, Sourav Sarkar and Jan-e Alam}
\medskip
\affiliation{Theoretical Physics Division, 
Variable Energy Cyclotron Centre, 1/AF, Bidhan Nagar, 
Kolkata - 700064}

\date{\today}
\begin{abstract}
The drag and
diffusion coefficients of a hot hadronic medium consisting of
pions, kaons and eta using open beauty mesons as a probe have been evaluated.
The interaction of the probe with the  hadronic matter 
has been treated in the framework of chiral perturbation theory.
It is observed that the magnitude of both the transport coefficients  
are significant, indicating substantial amount of interaction of the
heavy mesons with the thermal bath. The 
results may have significant impact on the  experimental 
observables like the suppression of  single electron spectra originating 
from the decays of heavy mesons produced 
in nuclear collisions at RHIC and LHC energies.   
\end{abstract}

\pacs{12.38.Mh,25.75.-q,24.85.+p,25.75.Nq}
\maketitle

\section{Introduction}
The suppression of the transverse momentum ($p_T$) distribution of 
hadrons produced in nucleus-nucleus  relative to
(binary scaled) proton-proton interactions at the Relativistic
Heavy Ion Collider (RHIC)~\cite{npa2005}  has been used as a tool
to understand the properties of matter formed in such collisions.
The large value of the elliptic flow of hadrons  measured
at RHIC along with the suppression of the high $p_T$ hadrons 
mentioned above indicate that the matter might have been formed 
in the partonic phase with liquid like properties characterized by
low value of shear viscosity ($\eta$) to entropy density
($s$) ratio, $\eta/s$ with a lower bound of $\eta/s\,\sim 1/4\pi$~\cite{KSS}. 

In addition to elliptic flow ($v_2$) and nuclear suppression 
($R_{\mathrm AA}$)  of light hadrons these quantities have
also been measured  for the single electron spectra 
originating from the decays of the open charm and beauty
mesons produced at RHIC collisions~\cite{raaexpt,v2expt}.
The advantages with heavy mesons are two-fold. Firstly, they contain
either a charm or a beauty quark which is produced 
very early and hence can witness the evolution of the
partonic matter since its inception until it reverts to hadronic 
matter through phase transition/cross over~\cite{lqcd} and secondly, 
the heavy quarks do not decide the bulk properties of the 
latter. Therefore, charm and beauty quarks are considered to be
efficient probes 
for the characterization of the partonic phase.  
In most of the earlier works
~\cite{hvh1,hvh2,ko,adil,gossiaux,wicks,das,hirano,alberico,
MBAD,ars,teaney,Das3} aimed at extracting the properties 
of quark gluon plasma (QGP) by analyzing the
$R_{\mathrm AA}$ and $v_2$ of heavy flavours the role of 
the hadronic matter was ignored. 
However, for the characterization of QGP  the 
interactions of heavy flavours with hadronic matter
should be taken into consideration and the
effects of  hadrons must be subtracted out from the observables.  
Though a large amount of work has been done 
on the diffusion of heavy quarks in the QGP the diffusion of
heavy mesons in hadronic matter has received much
less attention so far. Recently the diffusion coefficient 
of $D$ meson has been  calculated  using heavy meson 
chiral perturbation theory~\cite{laine} and also by using the empirical 
elastic scattering amplitudes~\cite{MinHe} of $D$ mesons with thermal hadrons.
The $D$-hadron interactions also have been evaluated using Born amplitudes~\cite{Ghosh} 
and unitarized chiral effective $D\pi$ interactions~\cite{abreu}. 
It has been found that the contributions of $B$ meson to the
single electron spectra dominate over those from $D$ meson
for large transverse momentum, $p_T>5$ GeV~\cite{djordjevic} 
(see also~\cite{Rapp_new}).
Moreover, the future experiments are 
progressing toward precision measurement over a wide range of 
kinematical variables. In view of this the use of $B$ meson 
as a probe to extract the properties of matter at high
temperature  assumes importance.  

In the next section we discuss the formalism adopted to 
evaluate the drag and diffusion coefficients of the 
heavy flavoured mesons in a hadronic matter consists  of  
pions, kaons and eta. Results are presented in section III 
and section IV is dedicated to summary and discussions.

\section{Formalism}
In the present work the drag and diffusion coefficients 
of the $B$ meson propagating through
a hot hadronic matter are evaluated within the ambit of 
Heavy Meson Chiral
Perturbation Theory $(HM\chi PT)$ in LO, NLO and NNLO approximations. 
We also revisit the transport coefficients 
of $D$ meson in a similar theoretical framework.
We consider the elastic interaction 
of the $B$ meson with thermal  
pions, kaons and eta in the temperature ($T$) 
range $100-170$ MeV. %The inelastic interaction has been ignored because 
Detailed analysis of the experimental data on the hadronic yield  in heavy ion 
collisions show that the value of the temperature for the chemical freeze-out 
of the system produced at RHIC energies is about 170 MeV (see ~\cite{chfo}
for a review).
This indicates that the inelastic interactions which are responsible 
for the change in the number of hadrons become rarer for $T$ 
below $\sim 170$ MeV.
Thus the contributions of the inelastic collisions in
evaluating the drag and diffusion coefficients of the hadronic matter
probed by the heavy flavoured mesons can be ignored. 

The drag ($\gamma$) and diffusion ($B_0$)
coefficients of the heavy mesons
are evaluated using elastic interaction
with the thermal hadrons.
For the (generic) process, $B(p) + h(q) \rightarrow B(p^\prime) + h(q^\prime)$ 
($h$ stands for pion, kaon and eta),
the drag $\gamma$ can be calculated by using the following 
expression~\cite{BS}: % (see also ~\cite{DKS}):
\begin{equation}
\gamma=p_iA_i/p^2
\label{drag}
\end{equation}
where $A_i$ is given by 
\begin{eqnarray}
A_i&=&\frac{1}{2E_p}\int\frac{d^3q}{(2\pi)^3E_q×}\int\frac{d^3p^\prime}{(2\pi)^3E_p^\prime×}
\int \frac{d^3q^\prime}{(2\pi)^3E_q^\prime×}\nonumber\\ 
&&\times\frac{1}{g_B}
\sum  \overline{|M|^2} (2\pi)^4 \delta^4(p+q-p^\prime-q^\prime)f(q)\nonumber\\
&&(1+f(q^\prime))[(p-p^\prime)_i] \equiv \langle \langle
(p-p^\prime)\rangle \rangle
\label{eq1}
\end{eqnarray}
$g_B$ being the statistical degeneracy of the $B$ meson.
The factor $f(q)$ denotes the thermal phase space factor 
for the particle in the incident channel and $1+f(q^\prime)$
is the Bose enhanced final state phase space factor. 
From Eq.~\ref{eq1} it is clear that the drag coefficient is a
measure of the thermal average of the momentum transfer, $p-p^\prime$,
weighted by the interaction through the square of the invariant amplitude,
$\overline{\mid M\mid^2}$. 

The diffusion coefficient, $B_0$ can be defined as:
\begin{eqnarray}
B_0=\frac{1}{4}\left[\langle \langle p\prime^2 \rangle \rangle -
\frac{\langle \langle (p.p\prime)^2 \rangle \rangle }{p^2}\right]
\label{diffusion}
\end{eqnarray}

Both the drag and diffusion coefficients 
can be evaluated from a single expression:
\begin{eqnarray}
\langle \langle \Gamma(p)\rangle \rangle &=&\frac{1}{512\pi^4×}\frac{1}{E_p} 
\int_{0}^{\infty} \int_{-1}^{1}d(cos\theta_{cm})\nonumber\\
&\times&\int_0^{2\pi}d\phi_{cm}
\frac{q^2 dq d(cos\chi)}{E_q}
(1+f(q^\prime))\nonumber\\
&\times&\hat{f}(q)\frac{\lambda^{\frac{1}{2}}(s,m_p^2,m_q^2)}{\sqrt{s}} 
\frac{1}{g} \sum  \overline{|M|^2}{\cal{T}}(p\prime)\nonumber\\
\label{transport}
\end{eqnarray}
with an appropriate choice of ${\cal{T}}(p\prime)$.
In Eq.~\ref{transport} $\lambda(x,y,z)=x^2+y^2+z^2-2xy-2yz-2zx$,
is the triangular function.
%ccccccccccccccccccccccccccccccccccccccccccccccccccccccccc

We start our discussion on the determination of the scattering amplitudes
with the Lagrangian of Covariant Chiral Perturbation Theory $(C\chi PT)$
involving the heavy
$B$ (or $D$)  mesons~\cite{Geng} given by
%Covariant formalism of Chiral Perturbation Theory $(C\chi PT)$~\cite{Geng} as well as 
%from Heavy Meson Chiral Perturbation Theory $(HM\chi PT)$~\cite{Liu}.
\bea
{\cal L}_{C\chi PT}&=& \langle{\D_\mu P \D^\mu P^\dagger}\rangle  
-m_B^2\langle{P P^\dagger }\rangle \nonumber\\
&&- \langle{ \D_\mu P^{*\nu} \D^\mu P^{*\dagger}_\nu }\rangle
+ m_{B^*}^2\langle{P^{*\nu} P^{*\dagger}_\nu }\rangle  \nonumber\\
&&+ ig\langle{P^{*}_{\mu} u^\mu P^{\dagger}-P u^\mu P^{*\dagger}_\mu }\rangle +......
\label{eq:L_CXPT}
\eea
where the heavy-light pseudoscalar meson triplet $P=(B^{0}, B^{+},B^{+}_s)$, 
heavy-light vector meson triplet $P_\mu^*=(B^{*0}_\mu, B^{*+}_\mu,
B^{*+}_{s\mu})$  and $\langle...\rangle$ denotes trace in flavor space.
The covariant derivatives are defined as
$\D_\mu P_a = \del_\mu P_a - P_b\Gm^{ba}_\mu$ and
$\D^\mu P_a^\dag = \del^\mu P_a^\dag + \Gm_{ab}^\mu P_b^\dag$
with $a,b$ are the $SU(3)$ flavor indices.

The vector and axial-vector currents are respectively
given by $\Gm_\mu=\frac{1}{2}( u^\dagger\del_\mu u + u\del_\mu u^\dagger)$
and $u_\mu=i( u^\dagger\del_\mu u - u\del_\mu u^\dagger)$
where $ u = \exp(\frac{i\Phi}{2 F_0})$. The unitary matrix $\Phi$ collects the
Goldstone boson fields and is given by

$\Phi=\sqrt{2}\left(\begin{array}{ccc}
\frac{\pi^0}{\sqrt{2}}+\frac{\eta}{\sqrt{6}}
& \pi^{-} & K^{-} \\ \pi^{+}
& -\frac{\pi^0}{\sqrt{2}}+\frac{\eta}{\sqrt{6}}
& K^0 \\ K^+ & K^0 & -\frac{2\eta}{\sqrt{6}}\end{array}
\right)$. 

To lowest order in $\Phi$ the vector and axial-vector currents are:
\be
\Gm_\mu=\frac{1}{8 F_0^2} [\Phi , \del_\mu \Phi], ~~~~~u_\mu=-\frac{1}{F_0}
\del_\mu \Phi~.
\ee

From the first term (or kinetic part of the 
$P$-fields) of ${\cal L}_{C\chi PT}$ , 
the matrix elements for contact diagram
in terms of Mandelstam variables $(s,t,u)$ are obtained as

\bea
&&{M}_{B^+\pi^+}=-{M}_{B^+\pi^-}=-\frac{1}{4\F^2}(s-u) 
\nonumber\\
&&{M}_{B^+\pi^0}={M}_{B^+\eta}={M}_{B^+K^+}={M}_{B^+K^-}=0
\nonumber\\
&&{M}_{B^+\oK^0}=-{M}_{B^+K^0}=-\frac{1}{4F_{K}^2}(s-u) 
\label{M_Geng}
%\label{eq:M_Geng}
\eea
These can be represented in the isospin basis as 

\bea
&&{M}^{(3/2)}_{B\pi}=-\frac{1}{4\F^2}(s-u)~~~~{M}^{(1/2)}_{B\pi}=\frac{1}{2\F^2}(s-u)
\nonumber\\
&&{M}_{B\eta}=0~~~{M}^{(1)}_{BK}=0~~~{M}^{(0)}_{BK}=-\frac{1}{2F_{K}^{2}}(s-u)
\nonumber\\
&&{M}^{(1)}_{B\oK}=-{M}^{(0)}_{B\oK}=-\frac{1}{4F_{K}^2}(s-u) 
\label{M_Geng_iso}
\eea
where the isospin of the $B\Phi$ system appears in the superscript.
Denoting the threshold matrix elements by $T$, these are obtained from 
(\ref{M_Geng_iso}) and are given by
\bea
&&{T}^{(3/2)}_{B\pi}=-\frac{m_B m_{\pi}}{\F^2}~~~~{T}^{(1/2)}_{B\pi}=\frac{2 m_B m_{\pi}}{\F^2}
\nonumber\\
&&{T}_{B\eta}=0~~~{T}^{(1)}_{BK}=0~~~{T}^{(0)}_{BK}=-\frac{2 m_B m_K}{F_{K}^{2}}
\nonumber\\
&&{T}^{(1)}_{B\oK}=-{T}^{(0)}_{B\oK}=-\frac{m_B m_K}{F_{K}^2}
\label{T_Geng_iso}
\eea

One can reproduce these $T$-matrix elements in the isospin basis using
the lowest order $HM\chi PT$ Lagrangian for heavy mesons containing a
 heavy quark $Q$ and a light antiquark of flavor $a$ as given below
\cite{Manohar_Wise}
\bea
{\cal L}_{HM\chi PT}&=&-i~ tr_D(\bar H_{a}^{Q}v^\mu \del_\mu H_{a}^{Q})
\nonumber\\
&&-i~ tr_D(\bar H_{a}^{Q} v^\mu \Gamma^{ab}_\mu H_{b}^{Q})
\nonumber\\
&&+ \frac{g}{2}~ tr_D(\bar H_{a}^{Q} \gamma^\mu \gamma^5 u^{ab}_\mu H_{b}^{Q}) + ...
\label{eq:L_HMXPT}
\eea
where $H_{a}^{Q}=\frac{1+\slv}{2} (P^{*}_{a\mu} \gamma^\mu + iP_a \gamma^5) $ 
and $\bar H_{a}^{Q}=(P^{*\dagger}_{a\mu} \gamma^\mu + iP^{\dagger}_a 
\gamma^5)\frac{1+\slv}{2}  $ 
and $tr_D $ denotes trace in Dirac space.
In this formalism, since the factor $\sqrt{m_P}$ and $\sqrt{m_{P^*}}$
 have been absorbed into the $P_a$ and $P^{*}_{a\mu}$ fields, 
the threshold $T$-matrix element $(\tilde{T}^{P\Phi}_{th})$ now has the dimension 
of scattering length $a_P$ whereas in 
$C\chi PT$, we get a dimensionless $T$-matrix element $(T^{P\Phi}_{th})$.
The relation between these two $T$-matrix elements and the scattering length $a_P$ 
is given by
\be
T^{P\Phi}_{th}=m_P \tilde{T}^{P\Phi}_{th} = 8\pi (m_\Phi +m_P) a_P
\ee

The square of the isospin averaged  $T$-matrix element is given by
\be
\sum \overline{|{T}_{B\Phi}|^2}=\overline{|{T}_{B\pi}|^2} + \overline{|{T}_{BK}|^2} + \overline{|{T}_{B\oK}|^2}
\ee
where $\overline{|{T}_{B\pi}|^2}=\frac{1}{(2+4)}(2 |{T}^{(1/2)}_{B\pi}|^2+4|{T}^{(3/2)}_{B\pi}|^2)$ \\

and $\overline{|{T}_{BK/\oK}|^2}=\frac{1}{(1+3)}(|{T}^{(0)}_{BK/\oK}|^2+3|{T}^{(1)}_{BK/\oK}|^2)$.\\

\section{Results}
We  evaluate  the drag coefficients of the  
$B$-meson by using the momentum 
dependent and  momentum independent 
matrix elements  given by Eqs. ~\ref{M_Geng_iso}
and ~\ref{T_Geng_iso} respectively. 
The results are shown by the dot dashed
and dashed lines in Fig.~1. Inspired by the fact that 
the results for the two scenarios are not drastically 
different in the LO we proceed to  
evaluate the drag coefficient of heavy mesons 
by replacing $\sum \overline{|M|^2}$ by $\sum \overline{|{T}|^2}$ in NLO 
and NNLO also
where the $T$-matrix elements will be obtained from the scattering lengths.

\begin{figure}
\begin{center}
\includegraphics[scale=0.40]{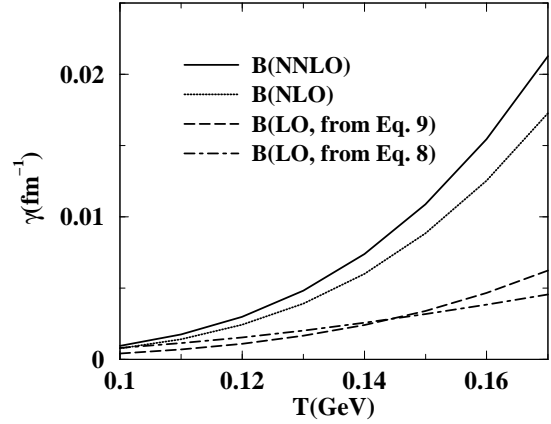}
\caption{The variation of drag coefficients with temperature
due to the interaction of  
$B$ mesons (of momentum $100$ MeV) 
with thermal pions, kaons and eta. 
The dot-dashed (dashed) line indicates the results obtained by using the  
matrix elements of Eq.~\ref{M_Geng_iso} ($T$ matrix of Eq.~\ref{T_Geng_iso}).
Solid (dotted) line indicates the results for NNLO (NLO) contribution \cite{Liu}.
}
\label{fig1}
\end{center}
\end{figure}

Liu {\it et al}~\cite{Liu} have obtained the $B\Phi$ scattering lengths (see
also ~\cite{Guo_new}) up to NNLO
in $HM\chi PT$ by using the coupling constant from recent unquenched lattice 
results \cite{Lat_g}. Using these NLO and NNLO results we estimate 
the isospin averaged drag coefficients of $B$ mesons. The results are depicted 
in Fig.~\ref{fig1}. 
The drag coefficient evaluated with NNLO matrix elements increases 
by $22\%$ compared to the NLO result at $T=170$ MeV. 

%cccccccccccccccccccccccccccccccccccccccccccccccccccccccc
%%%%%%%%%%%%%%%%%%%%%%%%%%%%%%%%%%%%%%%%%%%%%%%%%%%%%%%%%%%%
%%%%%%%%%%%%%%%%%%%%%%%%%%%%%%%%%%%%%%%%%%%%%%%%%%%%%%%%%%%%%%%%%%
We now focus on the temperature dependence of the drag coefficient
of $B$-mesons as shown in Fig.~\ref{fig1}. 
As mentioned before, $\gamma$ is the thermal average of the
square of the  momentum exchanged between the heavy mesons and the
bath particle weighted by the interaction strength 
through  the invariant amplitude of the process. 
Therefore, with the increase in temperature of the thermal bath
the kinetic energy of the hadrons increases.
Hence the hadrons gain the ability to transfer larger momentum during their
interaction with the $B$ mesons - resulting in
the increase of the drag coefficient.
This tendency is observed in Fig.~\ref{fig1} quite clearly.
The increase of drag with temperature is characteristic
of a gaseous system. 
In case of a liquid the
drag diminishes with $T$ (except for very few cases). 
In this case a significant part of the thermal
energy goes into making the attraction between the interacting
particles weaker and once this happens
the constituents move more freely resulting in
a smaller drag force. Therefore, the variation of the drag 
with temperature can be used to understand the nature of 
interaction of the fluid.

%%%%%%%%%%%%%%%%%%%%%%%%%%%%%%%%%%%%%%%%%%%%%%%%%%%%%%%%%%%%
\begin{figure}
\begin{center}
\includegraphics[scale=0.40]{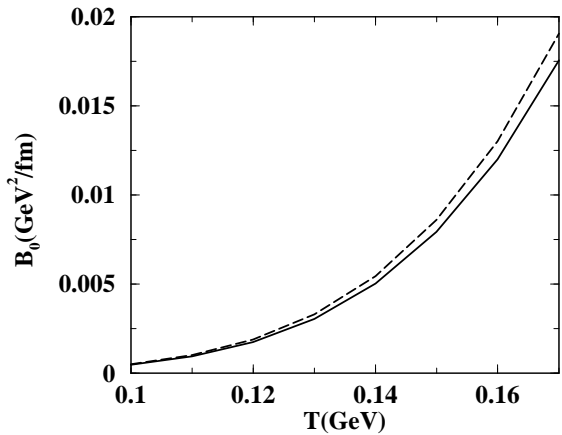}
\caption{Variation of diffusion co-efficient as a function of temperature.
The solid line indicates the variation 
of the diffusion coefficient with temperature
obtained from Eqs.~\ref{diffusion} and~\ref{transport}.
The momentum of the $B$ meson is taken as 100 MeV.
The dashed line stands for the diffusion coefficient
obtained from the Einstein relation (Eq.~\ref{Einstein}).
}
\label{fig2}
\end{center}
\end{figure}
%%%%%%%%%%%%%%%%%%%%%%%%%%%%%%%%%%%%%%%%%%%%%%%%%%%%%%%%%%%%%%%%%%
%%%%%%%%%%%%%%%%%%%%%%%%%%%%%%%%%%%%%%%%%%%%%%%%%%%%%%%%%%%%
\begin{figure}[ht]
\begin{center}
\includegraphics[scale=0.40]{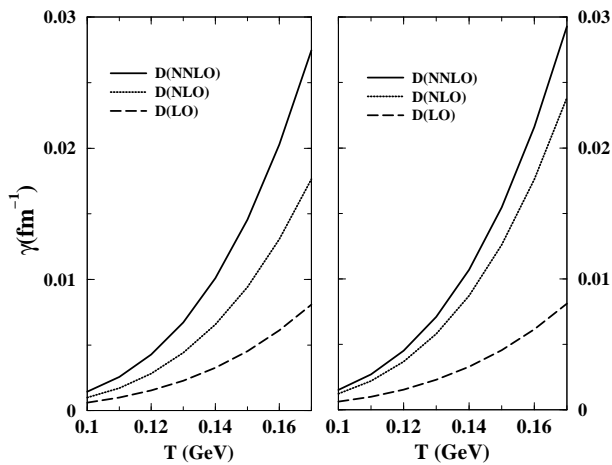}
\caption{The variation of drag coefficients of $D$ mesons with temperature
due to interaction with thermal pions, kaons and eta 
in LO, NLO and NNLO approximations for interactions of $D$ with thermal
hadrons taken from Ref.~\cite{Geng} (left panel) and~\cite{Liu} (right panel).
}
\label{fig3}
\end{center}
\end{figure}
%%%%%%%%%%%%%%%%%%%%%%%%%%%%%%%%%%%%%%%%%%%%%%%%%%%%%%%%%%%%%%%%%%

Since the diffusion coefficient involves the square of
the momentum transfer it is also expected to increase with
$T$. This is seen in Fig.~\ref{fig2}. 
%The diffusion coefficients 
%increases with $T$ because the average value of the
%square of the momentum transfer increases with $T$. 
The drag  and the diffusion coefficients are related 
through the Einstein  relation as: 
\be
B_0=M_B\gamma T.
\label{Einstein}
\ee
where $M_B$ is the mass of the $B$-meson. 
The temperature dependence of the diffusion coefficient 
evaluated by using Eqs.~\ref{transport} and the \ref{Einstein} are displayed  
in Fig.~\ref{fig2}. The  
difference between the results obtained from Eq.~\ref{transport} 
and the Einstein's relation is about $6-7\%$
at $T=170$ MeV for the $B$ meson momentum, $p=100$ MeV. 
This small difference 
illustrates the validity of the Einstein relation
in the low momentum (non-relativistic) domain.

The energy loss of a  $B$ meson moving through a 
hadronic system may be estimated from the relation
\be
-\frac{dE}{dx}=\gamma p~.
\label{dedx}
\ee
The magnitude of $\gamma$ obtained in the present calculation  
reveals that the $B$ mesons dissipate significant amount
of  energy in the medium. This might have crucial 
consequences on
quantities such as the nuclear suppression factor of single
electrons originating from the decays of heavy mesons.

We also evaluate the $D$ meson drag and diffusion coefficients  
using the interactions of $D$ mesons  with thermal
hadrons discussed  in  Refs.~\cite{Geng} 
and ~\cite{Liu} in LO, NLO and NNLO
approximations.  The results are displayed in Fig.~\ref{fig3}. 
In the LO approximation the drag is similar for both the cases. However, 
for NLO and NNLO, the drag co-efficient evaluated using the $T$-matrix elements
obtained from the
scattering lengths of Ref.~\cite{Liu} is slightly higher 
than that obtained from Ref.~\cite{Geng}. 

The drag of $D$ mesons in hot hadronic matter has recently
been studied by using  different approaches.
While empirical scattering cross sections were used in Ref.~\cite{MinHe},
the authors of Ref.~\cite{abreu} used unitarized chiral effective $D\pi$ 
interactions to evaluate the drag. 
We observe that the magnitude of the drag of $D$ meson obtained 
in the present work is similar to that  obtained 
in Refs.~\cite{abreu} and \cite{MinHe}. The smaller value in the present case
is due to the lower
values of the $D$ meson-hadron cross sections.

\section{Summary and discussions}
In summary we have evaluated the drag and diffusion coefficients
of open beauty mesons interacting with a hadronic background composed of 
pions, kaons and eta. It is found  that the values of both the 
transport coefficients increase with temperature.
The magnitude of the drag coefficient
of the $B$ meson indicates 
that while evaluating the  suppression  of the high $p_T$ 
single electrons originating from the decays of $B$ mesons 
the effects of hadrons should be taken into account. 
Within the same formalism, the 
transport coefficients of the $D$ meson has been calculated. The $D$ 
meson drag coefficient is found to be lower than the
values obtained in Refs. ~\cite{MinHe,abreu}.

Some  comments on the effects of the exclusions of inelastic channels and
non-perturbative processes on the drag and diffusion coefficients 
are in  order here.  
The  inelastic channels do contribute to the scattering matrix through
coupled channels via loops in the unitarisation procedure.
However, in a thermal background such as in a heavy ion collision
the scattering amplitude is weighted by phase space factors which
essentially control the rate of reactions.
The fact that heavy ion collisions undergo chemical
freeze-out at about 170 MeV means that number changing reactions
will certainly be inhibited below this temperature.
The presence of resonances makes the
evaluation of scattering amplitudes a non-perturbative problem.
Unitarisation of scattering amplitudes, preferably with loops evaluated
with thermal propagators will certainly improve the reliability of our
results at higher energies. However, 
in the absence of any information about B* mesons
so far, this exercise will however be far less constrained than the charm sector
where the masses and widths of the excited states are known.

{\bf Acknowledgment:} 

SKD and JA are partially  supported by DAE-BRNS project Sanction No.  2005/21/5-BRNS/2455.

\end{document}